\documentclass[traditabstract]{aa} % for the abstract without structuration 
\usepackage{graphicx}
\usepackage{natbib}
\usepackage{txfonts}
\usepackage{ulem}
\newcommand{\vhbb}{$\Delta V_{\rm HB}^{\rm Bump}\,$}
\newcommand{\vtob}{$\Delta V_{\rm TO}^{\rm Bump}\,$}
\newcommand{\vacstob}{$\Delta M_{F606W}^{\rm TO-Bump}\,$}
\newcommand{\vacshbb}{$\Delta M_{F606W}^{\rm HB-Bump}\,$}

\begin{document}
   \title{On the magnitude difference between the main sequence turn off and the red giant branch bump in Galactic globular clusters}
\titlerunning{On the magnitude difference between the turn off and the RGB bump}

   \author{S. Cassisi\inst{1}
          \and
          A. Mar{\'i}n-Franch \inst{2,3}
          \and
          M. Salaris\inst{4}
          \and
          A. Aparicio\inst{5, 6}
          \and
          M. Monelli\inst{6, 5}
          \and
          A. Pietrinferni\inst{1}
          }

   \institute{INAF - Osservatorio Astronomico di Teramo, Via M. Maggini, 64100 Teramo, Italy\\
              \email{cassisi,pietrinferni@oa-teramo.inaf.it}
         \and
            Centro de Estudios de Fisica del Cosmos de Aragon (CEFCA), E-44001 Teruel, Spain\\
            \email{amarin@cefca.es}
            \and
             Departamento de Astrofisica, Universidad Complutense de Madrid, E-28040 Madrid, Spain\\
           \and
     Astrophysics Research Institute, Liverpool John Moores University, Twelve Quays House, Birkenhead CH41 1LD, UK\\
      \email{ms@astro.livjm.ac.uk}
        \and
Departamento de Astrof\'{i}sica, Universidad de La Laguna,  Tenerife, Spain.
   \and
   Instituto de Astrof\'{i}sica de Canarias, E-38200 La Laguna, Tenerife, Spain\\
   \email{aparicio,monelli@iac.es}
               }

   \date{Received ; accepted}

% \abstract{}{}{}{}{} 
% 5 {} token are mandatory
 
  \abstract{We present new measurements of the magnitude of the main sequence turn off and the red giant branch bump in the luminosity function 
   of a sample of Galactic globular clusters with updated estimates of [Fe/H] and [$\alpha$/Fe], employing
   photometric data collected with the Advanced Camera for Survey on board the HST.
   We compare measured and predicted values of the magnitude difference between these two features, a rarely employed diagnostic 
   of the internal structure of low-mass stars at the beginning of their red giant evolution.
   Our analysis discloses a clear discrepancy between theory and observations, the theoretical red giant branch bump magnitudes being too bright by 
   on average $\sim0.2$~mag. This corroborates results from the more widely studied magnitude difference between 
   horizontal branch and red giant bump, avoiding the well known problems associated to the determination of the horizontal branch level 
   from colour magnitude diagrams, and to uncertainties in the luminosity of horizontal branch stellar models.
   We briefly discuss several potential solutions of this discrepancy.
}

   \keywords{stars: evolution $-$ stars: Population II $-$ Galaxy: globular clusters: general}

   \maketitle
%
%________________________________________________________________

\section{Introduction}

Several features of colour-magnitude diagrams (CMDs) and luminosity functions (LFs) of Galactic Globular Clusters (GCs) 
can be employed to test the accuracy of low-mass, metal-poor stellar models \citep[see, e.g.][]{rfs88}.
The bump appearing in the GC Red Giant Branch (RGB) LF is one of these important benchmarks.  
It is produced by the encounter of the H-burning shell with the 
H-abundance discontinuity left over by the outer convection at its maximum depth \citep{tho, ibe} reached during the first dredge-up.  
The sharp increase of the H-abundance causes a sudden decrease of the mean molecular weight ($\mu$), 
that affects the efficiency of the H-burning shell, proportional to a high power of $\mu$ \citep[see][]{kipwei,salcas}. 
This occurrence causes a temporary drop of the surface luminosity, before it starts to increase again.  
As a consequence, a low-mass RGB star crosses the same luminosity interval three times, and a 
bump (over-density) appears in the RGB differential LF (star counts per magnitude bin) of GCs 
\citep[for a detailed discussion we refer to][]{salaris02}. 
Given that the RGB-bump brightness depends on the maximum depth attained by the convective 
envelope, and the chemical profile above the advancing H-burning shell,  
the comparison between predicted and observed luminosity of the RGB-bump, 
provides valuable information about the internal structure of low-mass stars 
at the beginning of their RGB evolution.

Since its first detection in the LF of NGC104 \citep[47Tuc --][]{kdd} the RGB bump has been the 
subject of several theoretical and observational investigations \citep{fusipecci90, 
cassisi97, alves99, zoccali99, bono01, riello03, bjork06, dicecco10}. Thanks to these works, we have now
accurate measurements of its brightness in many GCs 
as well as in Local Group dwarf galaxies \citep[see][and references therein]{monelli10}.

The parameter routinely adopted to  
compare observations with theory 
is the quantity $\Delta V_{\rm HB}^{\rm Bump}= V_{Bump}-V_{HB}$, that is, the V-magnitude 
(or bandpasses similar to Johnson V)  
difference between the RGB-bump and the horizontal branch (HB) at the RR Lyrae 
instability strip level \citep{fusipecci90, cassisi97}. This has the advantage of being formally 
independent of distance and reddening, and not affected
by any uncertainty in the zero point of the photometry.
The most recent comparisons between \vhbb models and observations \citep[see, e.g., Fig.~10 in][]{dicecco10} 
seem to confirm a discrepancy (at the level of $\sim$0.20~mag or possibly more) 
for GCs with total metallicity [M/H] below $\sim -$1.5, in the sense that the 
predicted RGB-bump luminosity is too high. The quantitative estimate of the 
discrepancy depends on the adopted metallicity scale.  
At the upper end of the GC metallicity range, the existence of a discrepancy depends 
on the adopted metallicity scale.

One drawback of using \vhbb as diagnostic of the RGB-bump luminosity, is that 
uncertainties in the placement of the observed HB level for GCs with blue HB morphologies, 
and in theoretical predictions of the HB luminosity \citep[i.e., due to uncertainties 
in the calculations of the He-core mass at the He-flash, see e.g.][]{cas10}, hamper the interpretation 
of discrepancies between theory and observations. 

An alternative avenue explored in this paper is offered by measuring the 
magnitude difference between the Main Sequence (MS) Turn-Off (TO) and the RGB-bump brightness \vtob$= V_{TO}-V_{bump}$, that 
bypasses the HB.
Observationally, an accurate estimate of the TO brightness requires both very high quality photometric datasets, and a detailed analysis
of the uncertainty associated with the presence of binary stars. 
To the best of our knowledge, so far only \citet{caputo02} 
and \citet{meissner06} have studied the \vtob parameter. 
\citet{caputo02} used \vtob in combination with $\Delta V_{\rm HB}^{\rm TO}= V_{TO}-V_{HB}$ to 
investigate the metallicity scale of a large sample of galactic GCs, but did not attempt to assess the level of 
agreement between predicted and observed \vtob values. More recently, \citet{meissner06} used the \vtob together 
with other CMD age indicators, to check their mutual self-consistency. 
As a result, they found that the GC ages estimated from \vtob were younger by about 2~Gyr, in comparison with estimates based 
on the $\Delta V_{\rm HB}^{\rm TO}$ parameter. This occurrence was interpreted as an evidence that stellar models predict a too bright  
RGB-bump, by $\sim$0.2-0.3~mag.

We wish to reanalyze this issue employing new, accurate photometry of a large sample of GCs, that 
enabled us to determine both TO and RGB-bump magnitudes for 12 GCs, covering 
a large metallicity range. 
Our methodological approach is the following. We have first determined the apparent magnitudes of both TO and RGB-bump in our GC sample, 
and employed the cluster relative distances from a theoretical MS-fitting technique. As a second step, we have  
estimated individual cluster ages from the TO absolute magnitudes, obtained assuming the empirical MS-fitting distance to 
NGC6752 by \citet{grat03} as zero point of our relative distance scale. 
Another set of ages for each cluster is then determined from their observed \vtob, and compared with the TO ages. 
The outcome of this comparison constrains the level of agreement between predicted and observed RGB-bump luminosity, independently of the 
HB.

The plan of this paper is as follows: section~\ref{frame} presents briefly the observational dataset and 
the theoretical models adopted in our analysis; estimates and comparisons of TO and \vtob ages are described in section~\ref{comp}, followed 
by a final discussion.

\section{Observational and theoretical framework}\label{frame}

We have employed a subsample of F606W and F814W photometries from the ACS GC Survey Treasury Project \citep{ata2007}, 
and made use of the \vacstob parameter, that is the counterpart of \vtob in the F606W filter.
Details on the observations and data reduction, have been already discussed in \citet{ata2007} and \citet{jay08}. 

This database has been collected
mainly for the aim of investigating relative and absolute GCs ages, and the initial mass function of GC stars. 
The observational strategy was optimized to obtain accurate photometry of the faint portion of the CMD, and 
for many clusters the bright RGB photometry is saturated. 
In other clusters - belonging mainly to the metal poor tail of the GC metallicity distribution - there are so 
few RGB stars, that the RGB-bump detection is impossible. As a consequence, the number of GCs 
with measurements of \vacstob is reduced compared to the total 
number of objects in the original database (64 clusters). 
In addition, we chose to restrict our analysis to those globulars with recent (re)determinations of both [Fe/H] 
\citep{carretta09} and {$\rm [\alpha/Fe]$} \citep{carretta10}. More in detail, the values of [Fe/H] are obtained 
from Table~A.1 in \cite{carretta09}, that displays [Fe/H] estimates for 95 GCs, obtained transforming (and averaging) previous 
determinations onto the homogeneous scale set by high-resolution spectroscopic measurements on 19 clusters, reported 
in Table~1 of the same paper. This metallicity scale turns out to be very close to the \cite{zw} estimates.
The errors displayed in the same table  
(that represent the 1$\sigma$ rms with respect to the metallicity scale set by the high-resolution spectroscopy) are added in quadrature 
to the average systematic errors in the high resolution estimates of the 19 reference clusters (reported also in Table~1 of the same paper).
As for {$\rm [\alpha/Fe]$}, we assign a typical 0.10~dex uncertainty. 

Due to these additional constraints on the metallicity estimates, 
our sample is reduced to 11 objects, to which we added the NGC6341 (M92), using 
HST archive data (program 9453, PI T. Brown). For this latter cluster three images in each band were used, 
with exposure times of 0.5, 5, and 90~s in the $F606W$ band, and 0.5, 6, 
and 100~s in $F814W$. The photometry has been reduced with the 
DAOPHOT/ALLFRAME package \citep{alf}, and calibrated to the VEGAMAG
system following \citet{sirianni}. With the inclusion of M92 we are able to cover approximately  
the whole metallicity range of Galactic GCs. 

\begin{figure}
\centering
\includegraphics[width=8.6cm]{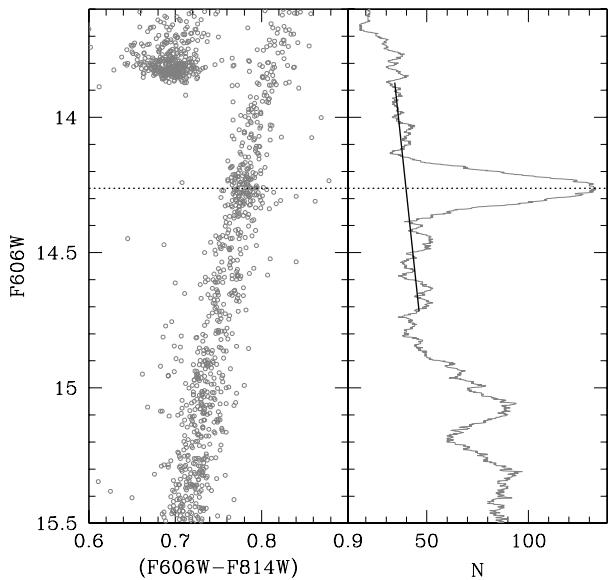}
\caption{{\sl Left panel}: CMD of NGC~104. {\sl Right panel}: the LF around the bump region. The solid line 
shows the continuum used for the bump determination (see text for more details), 
while the dotted line marks the bump location.}
\label{bumpFig}
\end{figure}

Measurements of the apparent TO magnitude and associated uncertainties are from \citet{amarin09}. In brief, these 
authors employed a  MS-fitting technique to determine the relative distance moduli between the reference cluster NGC6752 and all 
other clusters 
in our sample. We adopt here a zero point for these relative distances set by the empirical MS-fitting distance to NGC6752 
determined by \citet{grat03}\footnote{The empirical MS-fitting distance to NGC6752  
derived in \citet{grat03} was obtained by fitting the observed globular cluster mean locus to 
the colour magnitude diagram location of local subdwarfs with the same [M/H] of the cluster.
Although both [Fe/H] and [$\alpha$/Fe] values employed by \citet{grat03} are different 
from the values reported in Table~1 for this cluster, the total metallicity [M/H] turns out to be same within 0.01~dex.}.
The final errors on the absolute magnitudes $M^{TO}_{F606W}$ reported in Table~\ref{tabgc} 
have been obtained by adding in quadrature the 
errors on the determination of the apparent magnitudes, errors on the relative cluster distances \citep{amarin09}, and the error on NGC6752 
distance \citep{grat03}. 
As for the determination of the RGB-bump level, we have employed the following method, 
illustrated in Fig.~\ref{bumpFig} for the case of NGC~104.
For each cluster, the LF around the bump region has been determined using a bin size equal to 0.1~mag.  
A linear fit (see right panel of Fig.~\ref{bumpFig}) is then performed around the bump -- 
excluding the bump feature -- to obtain what we denote as the LF 'continuum' 
(marked as a solid black line in the right panel of Fig.~\ref{bumpFig}). Finally, the apparent magnitude of the bump (dotted line) 
is determined at the maximum of the continuum-subtracted LF, and its uncertainty is computed as $\sigma/\sqrt{N_{stars}-1}$, 
with $N_{stars}$ and $\sigma$ denoting the number of bump stars and their standard deviation around the bump luminosity, respectively. Results 
of these measurements as well as [Fe/H] spectroscopic estimates and the global metallicity [M/H] obtained from the measurements of 
[Fe/H] and [$\rm \alpha$/Fe] are reported in Table~\ref{tabgc}. 

Our theoretical analysis makes use of the $\alpha$-enhanced BaSTI\footnote{The BaSTI stellar evolution library is available at the 
following URL: http://www.oa-teramo.inaf.it/BASTI.} stellar model library \citep{pietrinferni06}
that employ transformations to the ACS/HST photometric system by \cite{bedin05}.
From this extended set of isochrones we have obtained the theoretical estimates of the TO, 
RGB-bump brightness, and \vacstob as a function of age and [M/H], that are displayed in Fig.~\ref{deltatobump}  
\footnote{Our analysis relies entirely on the theoretical predictions from the 
BaSTI models. Although a good agreement does exist among several different stellar model libraries 
concerning the trend with age of the evolutionary features discussed in this paper, some 
small, marginal differences can still be present.}.

It is interesting to investigate the sensitivity of the \vacstob parameter to the cluster age $t$,  
in comparison with \vacshbb, i.e. the equivalent of the traditional \vhbb.
In the age range typical of GCs, the HB is practically unaffected by the exact value of $t$, 
while the RGB-bump becomes fainter with increasing $t$. 
As a result \vacshbb increases with age by $\sim 0.03$ mag/Gyr. 
In case of \vacstob, both TO and RGB-bump become fainter with increasing age, but the effect of 
changing $t$ is larger on the TO luminosity. Overall 
\vacstob increases with age by $\sim 0.08$ mag/Gyr at [M/H]=$-2.0$, and by $\sim 0.04$ mag/Gyr at [M/H]=$-0.5$.

\begin{table*}
\caption{The metal content, TO, RGB bump brightness, and age estimates from both TO and \vacstob for the selected sample of GCs.}
\centering
\label{tabgc}
\begin{tabular}{ccccccccc}
\hline\hline
{Cluster}  & {[Fe/H]} & {[$\alpha$/Fe]} & {[M/H]\tablefootmark{a}} & {$m_{F606W}(TO)$} & {$M_{F606W}(TO)$} &  
{$m_{F606W}(Bump)$} & {Age(MSTO)\tablefootmark{b}} & {Age(\vacstob)\tablefootmark{c}} \\
\hline

 NGC0104  & -0.76 & 0.42 & -0.45$\pm$0.11 & 17.53$\pm$0.07 & 4.27$\pm$0.11 & 14.26$\pm$0.004 & 13.98$\pm$2.11 &  9.66$\pm$3.55\\       
 NGC0362  & -1.30 & 0.30 & -1.09$\pm$0.12 & 18.74$\pm$0.05 & 3.95$\pm$0.09 & 15.17$\pm$0.007 & 11.85$\pm$1.62 &  7.02$\pm$1.18\\       
 NGC1851  & -1.18 & 0.38 & -0.90$\pm$0.14 & 19.40$\pm$0.05 & 3.93$\pm$0.09 & 15.84$\pm$0.008 & 10.88$\pm$1.70 &  7.95$\pm$2.17\\       
 NGC5904  & -1.33 & 0.38 & -1.05$\pm$0.11 & 18.37$\pm$0.03 & 3.97$\pm$0.09 & 14.73$\pm$0.008 & 11.95$\pm$1.53 &  7.75$\pm$1.19\\       
 NGC6093  & -1.75 & 0.24 & -1.58$\pm$0.14 & 19.67$\pm$0.05 & 4.01$\pm$0.09 & 15.72$\pm$0.010 & 14.69$\pm$1.93 &  8.53$\pm$1.74\\       
 NGC6218  & -1.33 & 0.41 & -1.03$\pm$0.11 & 18.18$\pm$0.04 & 4.13$\pm$0.09 & 14.49$\pm$0.013 & 14.12$\pm$1.99 &  8.60$\pm$1.86\\       
 NGC6254  & -1.57 & 0.37 & -1.30$\pm$0.11 & 18.36$\pm$0.04 & 4.00$\pm$0.09 & 14.47$\pm$0.012 & 13.32$\pm$1.80 & 10.38$\pm$1.91\\       
 NGC6341  & -2.35 & 0.46 & -2.01$\pm$0.12 & 18.60$\pm$0.04 & 3.86$\pm$0.09 & 14.49$\pm$0.011 & 14.23$\pm$1.77 &  8.79$\pm$1.07\\       
 NGC6541  & -1.82 & 0.43 & -1.50$\pm$0.14 & 18.70$\pm$0.04 & 4.02$\pm$0.09 & 14.76$\pm$0.010 & 14.51$\pm$1.97 &  9.38$\pm$1.87\\       
 NGC6637  & -0.59 & 0.31 & -0.37$\pm$0.13 & 19.40$\pm$0.07 & 4.27$\pm$0.11 & 16.10$\pm$0.008 & 13.69$\pm$2.26 & 12.57$\pm$3.91\\       
 NGC6723  & -1.10 & 0.50 & -0.72$\pm$0.13 & 18.90$\pm$0.05 & 4.21$\pm$0.09 & 15.34$\pm$0.010 & 14.12$\pm$2.15 & 10.73$\pm$3.61\\     
 NGC6752  & -1.55 & 0.43 & -1.23$\pm$0.11 & 17.26$\pm$0.03 & 4.02$\pm$0.09 & 13.39$\pm$0.012 & 13.28$\pm$1.75 & 10.58$\pm$1.76\\    
\hline 
\end{tabular}
\tablefoot{
\tablefoottext{a}{The cluster global metallicity obtained by combining the [Fe/H] estimates from \cite{carretta09} and the average  
$\alpha-$elements enhancement from \cite{carretta10} according to Eq.~3 in \cite{scs93}. The associated error bar is obtained 
by propagating the errors in [Fe/H] and [$\alpha$/Fe] discussed in Sect.~2. accordingly.}
\tablefoottext{b}{The cluster age in Gyr estimated by using the TO magnitude and the theoretical calibration shown in the upper panel of 
Fig.~\ref{deltatobump}.}
\tablefoottext{c}{The cluster age in Gyr estimated by using the \vacstob parameter and the theoretical calibration shown in the lower panel of 
Fig.~\ref{deltatobump}.}}
\end{table*}

\section{Comparison between theory and observations}\label{comp}

\begin{figure}
\centering
\includegraphics[width=9.5cm]{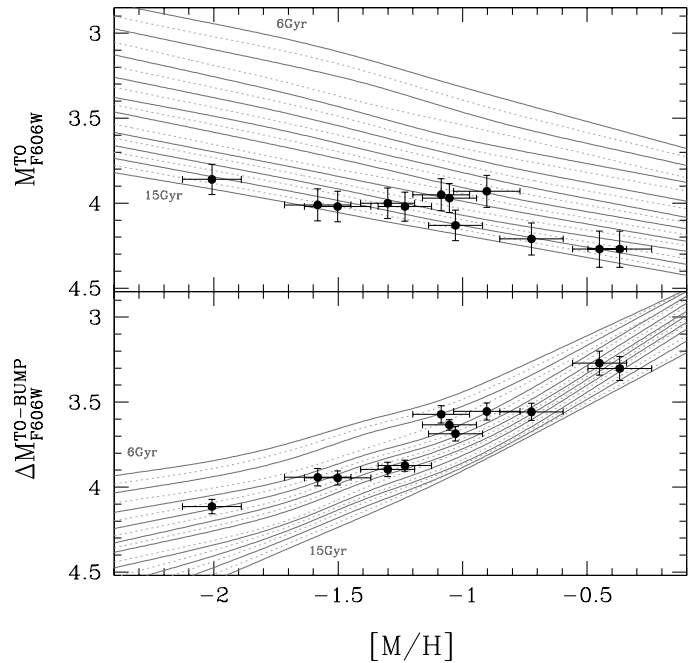}
\caption{{\sl Upper panel}: Absolute $M_{F606W}$ magnitude of the TO as a function of [M/H] for our GC sample (filled circles). 
Solid and dashed lines display the theoretical calibration from BaSTI $\alpha-$enhanced isochrones for ages between 
6 and 15~Gyr, in steps of 0.5~Gyr. 
{\sl Bottom panel}: As the upper panel but for \vacstob.}
\label{deltatobump}
\end{figure}

The values of \vacstob measured in our sample of 12 clusters are displayed in the lower 
panel of Fig.~\ref{deltatobump}, over-imposed to the theoretical 
calibration as a function of [M/H] and age. 
The cluster ages necessary to match the observed \vacstob appear generally younger 
than standard GC ages (of the order of 12-14~Gyr). 
The \vacstob ages and associated errors are reported in Table~\ref{tabgc}, as obtained by interpolation 
amongst the theoretical values. A conservative estimate of the associated error has been 
obtained by considering the rectangle defined in the \vacstob$-$[M/H] plane by  the the uncertainties in both 
\vacstob and [M/H]. The error in the age estimate has been then determined 
from the ages of the ``youngest'' and ``oldest'' corners of the rectangle.
The same approach has been followed also to estimate the uncertainties in the ages from the absolute TO magnitude.

The upper panel of Fig.~\ref{deltatobump} displays a comparison between the theoretical calibration of the TO absolute 
magnitude (as a function of age and [M/H]) and the cluster TO absolute magnitudes $M^{TO}_{F606W}$ (also reported in Table~\ref{tabgc}). 
A visual comparison of the upper and lower panels of Fig.~\ref{deltatobump} confirms that ages from \vacstob tend to be 
systematically lower than TO ages.

\begin{figure}
\centering
\includegraphics[width=9.0cm]{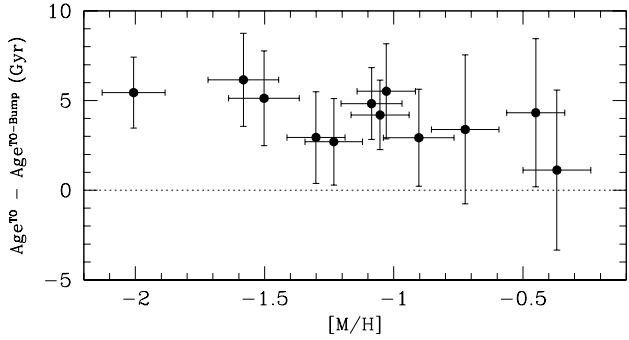}
\caption{Difference between the cluster ages inferred from the TO absolute magnitude and from \vacstob.}
\label{dage}
\end{figure}

\begin{figure}
\centering
\includegraphics[width=9.0cm]{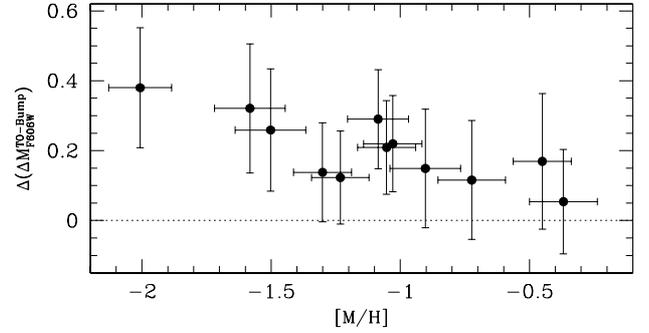}
\caption{Difference between the values of \vacstob expected from the cluster TO ages, and the measured values, 
as a function of [M/H].}
\label{difteo}
\end{figure}

Figure~\ref{dage} displays the difference between TO and \vacstob ages as a function of [M/H]. 
All points are systematically shifted to positive values of the age difference, and for about half of the clusters in the sample 
the difference is significant at the 2$\sigma$ level or more.
Another way to look at this discrepancy is to display the difference between the values of \vacstob expected from the 
cluster TO ages, and the measured values, as shown in  Fig.~\ref{difteo}. 
Conservative error bars on these $\Delta$(\vacstob) values have been 
obtained applying a procedure analogous to the one followed to determine the errors on \vacstob ages. 
Figure~\ref{difteo} shows very clearly that the expected \vacstob values are systematically larger 
(as can be also inferred from  Fig.~\ref{deltatobump}) than observed. 
The mean value of $\Delta$(\vacstob) is equal to 0.20~mag, with a 1$\sigma$ dispersion of $\pm$0.1~mag.
A linear fit that takes into account the errors on both $\Delta$(\vacstob) and 
[M/H] \citep[using the routine fitexy in][]{numrec} provides a slope d$\Delta$(\vacstob)/d[M/H]=$-0.16\pm0.12$,  
that is not significantly different from zero. Neglecting the more 
discrepant cluster with [M/H]$\sim-$2.0 (NGC6341) leaves the mean value of $\Delta$(\vacstob) almost unchanged 
(0.19~mag, with a 1$\sigma$ dispersion of $\pm$0.08~mag), whilst the slope of the linear fit is again not statistically 
significant (d$\Delta$(\vacstob)/d[M/H]=$-0.14\pm0.14$).

\section{Discussion}

The main result of our analysis is summarized by 
Fig.~\ref{difteo}, discussed in the previous section. 
The values of \vacstob predicted by theoretical models 
for cluster ages estimated from the TO absolute magnitudes, are larger than observed. 
Given that the observed TO magnitude is by definition matched by the theoretical isochrones to determine the TO age, 
this discrepancy implies that the absolute magnitude of the 
RGB-bump in the models is too bright. 

An extension of this type of analysis 
to a larger, homogeneous sample of GC photometries is obviously desirable; hovewer 
our results based on a limited sample of clusters provide already 
clear evidence of a real \lq{over-luminosity}\rq\ of the predicted absolute 
magnitude of the RGB-bump, irrespective of problems with HB 
modelling and placement of the reference HB level in clusters with only blue HB stars. 

The simplest explanation for this discrepancy envisages a systematic underestimate of the 
cluster metallicities by $\sim$0.2~dex. An higher [M/H] would    
eliminate the discrepancy, because it causes a lower TO age and a lower theoretical 
RGB-bump brightness for each cluster. There is of course no indication that the metallicity scale we adopted 
is affected by this type of systematics, but this is a point to be considered. 

In the following we expand our discussion to see how improvements in the 
micro- (e.g. opacities, nuclear reaction rates) and macro-physics (e.g., element transport meachanisms) 
employed in stellar evolution calculations, and the recently extablished presence of multiple stellar populations with 
varying chemical patterns in individual GCs, can account for this discrepancy.

\subsection{Improved micro- and macro-physics}

A straightforward explanation for the discrepancy highlighted in Fig.~\ref{difteo} could be an underestimate of the radiative opacity 
at temperatures around 
a few $10^6$~K -- typical temperatures at the lower boundary of the convective envelope. Higher opacities would shift the 
convection boundary  -- hence the H-abundance discontinuity  --  to deeper layers, causing a fainter RGB-bump.
However, this solution does not seem plausible,  
for the following reasons: i) radiative opacities in this temperature range should not be 
affected by an uncertainty larger than $\sim5$\% \citep[see, e.g.][]{guzik08} and this small change 
is not able to reconcile theory with observations; ii) the discrepancy theory-observations increases with decreasing [M/H], and 
it does not seem very likely that radiative opacities become less accurate when the metal content decreases. 

The isochrones employed to determine both the cluster ages from the TO brightness, and the theoretical values of \vacstob, do not account for 
the effect of atomic diffusion (including radiative levitation).
Although current spectroscopic observations of globular cluster stars show that atomic diffusion is at least partially 
inhibited by additional turbulence/mixing \citep[see, i.e.][]{korn07} -- induced for example by rotation 
\citep[see, i.e.,][]{egge10} -- we summarize here the effect on TO ages and 
\vacstob values in case of full efficiency.  
According to the results by \cite{vand02} and \cite{mich10} -- that expand upon previous studies by \cite{cds97, cassisi98} where the effect 
of radiative levitation was not considered -- atomic diffusion 
makes the RGB-bump magnitude brighter by 0.03-0.06~mag at fixed age, and also decreases the cluster TO ages 
by at most $\sim$1.5~Gyr for the lowest metallicity clusters.
The combined effect on \vacstob would decrease the discrepancy for the most metal poor cluster in our sample by $\sim$0.05~mag  at most. 
The effect becomes less significant with increasing metallicity.

On the other hand, the recent redetermination of the $^{14}N(p,\gamma)^{15}O$  reaction rate -- not included in our adopted models -- 
would increase the cluster ages by $\sim$ 1~Gyr, and at the same time make the RGB-bump brighter by $\sim$0.06~mag at fixed 
age \citep{weiss05, pietrinferni10}.
The net result would be an increase of the discrepancy by $\sim 0.10$~mag or more, that would move the 
mean value of $\Delta$(\vacstob) up to $\sim$0.30~mag.
Overall, the combined effect of the new $^{14}N(p,\gamma)^{15}O$ reaction rate and inclusion of atomic diffusion 
(plus radiative levitation) would exacerbate the discrepancy between theory and observations, 
that would become on average of the order of 0.25~mag.

Another possibility to mitigate the discrepancy is to include overshooting beyond the formal boundary 
of the convective envelope \citep[see, e.g.,][]{alongi91}. Calculations by \cite{cassisi02} show 
that the inclusion of convective overshooting decreases the RGB-bump brightness by ${\rm \sim 0.8 mag/H_P}$ (where $H_P$ denotes  
the local pressure scale height); the discrepancy between theory and observations would disappear with the inclusion of  
convective overshooting of the order of $\sim 0.25$ below the Schwarzschild boundary of the convective envelope. 

Besides overshooting from the convective boundary, \cite{cassisi02} 
have investigated also the effect on the RGB-bump shape and brightness, of a smoother chemical  discontinuity left over by 
the first dredge-up. A smoother chemical discontinuity could be produced, for example, by turbulent mixing counteracting 
the efficiency of atomic diffusion.  
\cite{cassisi02} results show that the bump luminosity decreases by $\sim$0.25 mag/${\rm H_p}$, 
where the smoothing length is expressed in units of the local pressure 
scale height. 
Given that smoothing the discontinuity alters also the shape of the RGB luminosity function in the bump region, this hypothesis 
is potentially testable. 
As estimated by \cite{cassisi02}, a sample of more than 120 RGB stars within $\pm$0.2~mag of the peak of the RGB-bump, 
and random photometric errors 
smaller than 0.03~mag can potentially disclose this effect in the RGB luminosity function. 

\subsection{The role of GC multipopulations}

A very important issue to be considered, is the effect on the cluster RGB-bump luminosity and TO ages of subpopulations with 
varying degrees of the CNONa anticorrelation and the -- likely -- associated increased He abundance, as observed in individual GCs 
\citep[see, e.g.,][for a review]{gcs04}. If the sum of the CNO abundance stays constant among all stars in a given cluster 
-- as observed, within the measurement errors -- the RGB-bump magnitude is affected only by the possible increase of helium. 
As shown by, e.g., \cite{cassisi97} and \cite{salaris06}, increasing the initial He abundance increases the bump brightness 
at fixed age and [Fe/H]. 
In a 'real' cluster the size of this effect depends on the exact amount of He-enhancement and the fraction of stars involved, 
but the main point is that this can only exacerbate the discrepancy displayed in Fig.~\ref{difteo}. 
As for the ages from the TO luminosity, one has to notice that within the individual clusters analyzed in this paper, there are no 
clear signs of large spreads of the initial He abundance, in terms of a split of the MS in the CMD. 
A reasonable upper limit to the He spread of 0.05 in mass fraction, 
would decrease the TO age by not more than $\sim$0.5~Gyr \citep[see, i.e.,][]{salaris06}.
As a conclusion, the effect of subpopulations with enhanced He within individual clusters in our sample would not 
solve the discrepancy highlighted by Fig.~\ref{difteo}.

Only NGC~1851 shows a clear split of the subgiant branch in our adopted CMD, whose origin is still debated 
\citep[see, e.g.,][]{cassisi08, carretta10b}.
The TO measurement has been obtained considering only the most populated SGB, that should harbour stars with a 
'standard' He and metal distribution \citep{cassisi08} so that also in this case our TO age estimates should be reliable.

\begin{acknowledgements}
We warmly thank our referee, Dr. A. Weiss, for his prompt report.
S.C. and A.P. acknowledge the partial financial support of INAF through the PRIN INAF 2009 (P.I.: R. Gratton) . 
This work was supported by the Science and Technology Ministry of the Kingdom of Spain 
(Consolider-Ingenio 2010 Program CSD 2006-00070, grants AYA2004-06343 and AYA2007-3E3507)
and by the IAC (grant 310394).
\end{acknowledgements}

\bibliographystyle{aa}

\end{document}